\documentclass[10pt,letterpaper,twocolumn]{article} 

\usepackage{ol2}
\usepackage[draft]{hyperref}
\usepackage{amsmath}

\begin{document}

\twocolumn[ 

\title{Detection of saturated absorption spectroscopy \\at high sensitivity with displaced crossovers}


\author{Jun Duan, Xianghui Qi, Xiaoji Zhou and Xuzong Chen$^*$}

\address{Institute of Quantum Electronics, School of Electronics Engineering
$\&$ Computer Science, \\ Peking University, Beijing 100871, P. R.
China}
\address{$^*$Corresponding author: xuzongchen@pku.edu.cn}

\begin{abstract}We present an unconventional experimental approach for detecting the saturated absorption
spectroscopy. Using this approach, crossover peaks are displaced,
leaving out peaks corresponding to atom's natural resonant
frequencies. Sensitivity of detection could also be enhanced.
Consequently the spectrum could reflect energy structure of atoms
more explicitly. Without harmful influence from crossovers, locking
range of the error signal is significantly increased and symmetry of
the dispersion line shape is perfectly reserved, so reliability of
frequency stabilization could be improved.\end{abstract}


 ] 

\noindent Saturated absorption spectroscopy (SAS) is a widely used
technique in laser frequency stabilization and atomic
physics\cite{The_Book_a, The_Book_b, The_Review, Cesium_SAS}.
Crossovers in such spectrum refer to extra peaks in the middle of
those indicating atom's natural resonant frequencies. They are
useful under certain circumstances\cite{Useful_Crossover} but can
also be undesired in other cases. For instance, in an optical pumped
cesium beam frequency standard, the pumping laser is stabilized on
the SAS of $F = 4 \to F'$ transitions of $^{133}$Cs D$_2$ hyperfine
structure. The transition of $F = 4 \to F' = 4$ is the most suitable
for obtaining an optimal pumping efficiency\cite{Rb_data}, but it is
near to the crossover of the $F = 4 \to F' = 3$ and the $F = 4 \to
F' = 5$ transitions, with a separation of about only
25MHz\cite{Wieman_Spectrum_a, Wieman_Spectrum_b}. This fact brings
negative influence for laser frequency stabilization. In general,
the smaller energy splitting between upper levels is, the more
obvious influence of crossovers is.

In this Letter, we demonstrate a method to avoid the influence of
crossovers on spectrum detection and frequency stabilization. We use
diode lasers at 780nm to excite $F = 2 \to F'$ transitions of
$^{85}$Rb D$_2$ hyperfine structure as an example. Crossovers are
displaced from their original position to prevent them from
overlapping with peaks which indicate atom's natural resonant
frequencies. At the same time more atoms contribute to the amplitude
of these peaks. So the spectrum could show energy structure of atom
more directly and distinctly in comparison with conventional SAS.
Furthermore, error signals obtained from such spectrum have larger
locking range and preserve better symmetry of the dispersion line
shape. Usage of these error signals could make laser frequency
stabilization more reliable.

\begin{figure}[htbp]
\centering
\includegraphics[width=8.3cm]{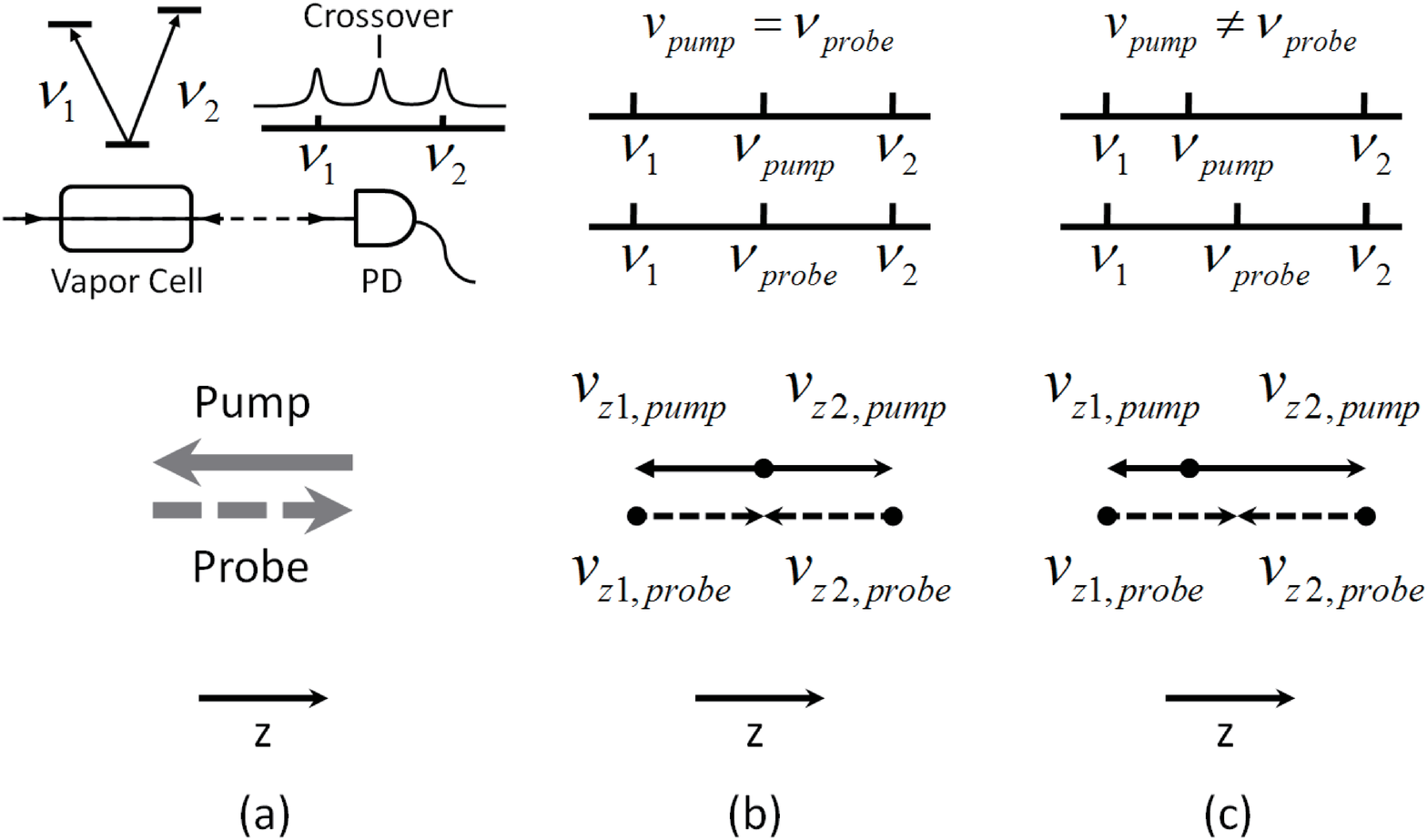}
\caption{How to move crossovers. ${\nu_{probe}} = ({\nu _1} + {\nu
_2})/2$. PD, photodiode. (a) Experimental setup. (b) Appearance of
crossover in conventional SAS. (c) Removal of crossover.}
\label{How_to_Remove_CO}
\end{figure}

To illustrate the displacement of crossovers in our experiment, we
can use a simple model of atom including only two excited states
corresponding to resonant frequencies $\nu_1$ and $\nu_2$
respectively (FIG. \ref{How_to_Remove_CO}(a)). The pump beam and the
probe beam pass through the absorbing sample (vapor cell) in
opposite directions, we take direction of the probe beam as
reference direction.

Peak of crossover appears right at the middle of $\nu_1$ and $\nu_2$
in the spectrum. Due to Doppler effect, when the probe beam is tuned
at frequency $\nu _{probe} = (\nu _1 + \nu _2)/2$ (FIG.
\ref{How_to_Remove_CO}(b), (c)), it employs atoms with velocity
class
\begin{equation}
{v_{z1,probe}} = \lambda (\nu _{probe} - {\nu _1}) = \lambda
\frac{{\Delta \nu }}{2} \label{v_z1,probe}
\end{equation}
to resonate with $\nu _1$ and atoms with velocity class
\begin{equation}
{v_{z2,probe}} = \lambda (\nu _{probe} - {\nu _2}) = - \lambda
\frac{{\Delta \nu }}{2} \label{v_z2,probe}
\end{equation}
to resonate with $\nu _2$, where $\Delta \nu  = {\nu _2} - {\nu
_1}$.

In a conventional SAS (FIG. \ref{How_to_Remove_CO}(b)), the pump
beam and the probe beam come from the same laser, so they always
have the same frequency. When the probe beam is tuned at $\nu
_{probe} = (\nu _1 + \nu _2)/2$, the pump beam is at this frequency,
too. It employs atoms with velocity class
\begin{equation}
{v_{z1,pump}} =  - \lambda ({\nu _{pump}} - {\nu _1}) =  - \lambda
\frac{{\Delta \nu }}{2} = {v _{z2,probe}} \label{v_z1,pump}
\end{equation}
to resonate with $\nu _1$ and atoms with velocity class
\begin{equation}
{v_{z2,pump}} = -\lambda (\nu _{pump} - {\nu _2}) = \lambda
\frac{{\Delta \nu }}{2} = {v _{z1,probe}} \label{v_z2,pump}
\end{equation}
to resonate with $\nu _2$. Atoms with velocity classes of $ \pm
\lambda \cdot \Delta \nu /2$ which are used by the probe beam are
also used by the pump beam, so a peak of crossover appears.

In our approach (FIG. \ref{How_to_Remove_CO}(c)), two diode lasers
are used to serve as the probe beam and the pump beam independently
and frequency of the pump beam is stabilized. They don't always have
the same frequency any more. For example, if $\nu _1 < \nu _{pump} <
(\nu _1 + \nu _2)/2 = \nu _{probe}$, the pump beam employs atoms
with velocity class
\begin{equation}
{v_{z1,pump}} =  - \lambda ({\nu _{pump}} - {\nu _1}) >  - \lambda
\frac{{\Delta \nu }}{2} = {v _{z2,probe}} \label{our_v_z1,pump}
\end{equation}
to resonate with $\nu _1$ and atoms with velocity class
\begin{equation}
{v_{z2,pump}} = -\lambda (\nu _{pump} - {\nu _2}) > \lambda
\frac{{\Delta \nu }}{2} = {v _{z1,probe}} \label{our_v_z2,pump}
\end{equation}
to resonate with $\nu _2$. Atoms used by the probe beam are not used
by the pump beam any more, this time when the probe beam is tuned at
$(\nu _1 + \nu _2)/2$, no peak will appear at this frequency. In
fact, there are still 3 saturated absorption peaks on the probe
spectrum, but they appear at different positions, and they are much
more distant from each other (detailed discussion will be presented
later). In other words, crossover is displaced.

\begin{figure}[htbp]
\centering
\includegraphics[width=8.3cm]{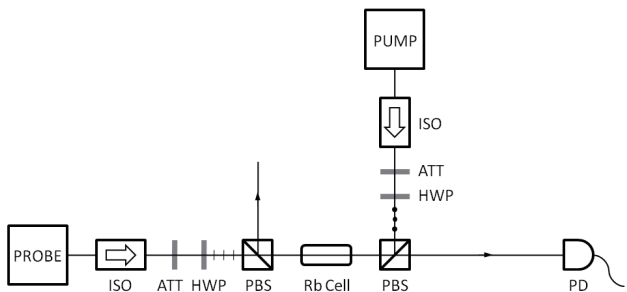}
\caption{Setup of our experiment. ISO, isolator; ATT, attenuator;
HWP, half-wave plate; PBS, polarizing beam splitter; PD,
photodiode.} \label{Experimental_Scheme}
\end{figure}

\begin{figure}[htbp]
\centering
\includegraphics[width=8.3cm]{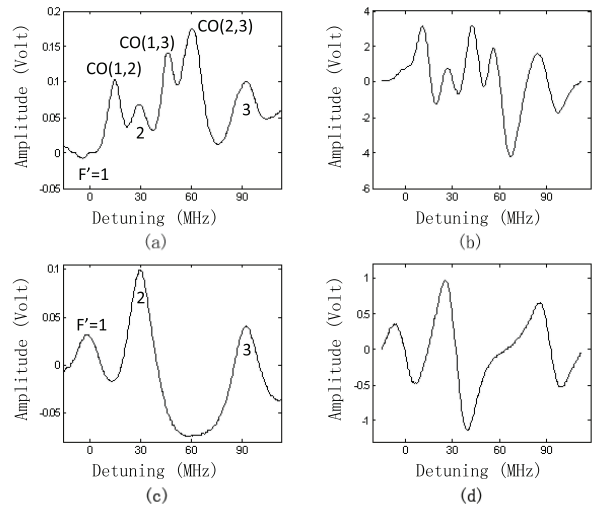}
\caption{Experimental results. (a) Conventional SAS. CO(1,2) is
short for crossover of $F = 2 \to F' = 1$ and $F = 2 \to F' = 2$.
(b) Corresponding error signals of conventional SAS in (a). (c)
Spectrum of our experiment. (d) Corresponding error signals of our
spectrum in(c).} \label{Experimental_Result}
\end{figure}

In our experiment (FIG. \ref{Experimental_Scheme}), frequency of the
pump beam is stabilized on the peak of $F = 2 \to F' = 3$ transition
of $^{85}$Rb D$_2$ hyperfine structure which is only little affected
by crossovers (the rightmost peak in FIG.
\ref{Experimental_Result}(a)) using a conventional SAS. Frequency of
the probe beam is scanned to get its absorption spectrum. Since the
pump laser is stabilized using conventional SAS, this method relies
on the stability of not only the probe laser, but also the pump
laser. Incident powers into the vapor cell are 0.4$\mu W$ for the
probe beam and 48$\mu W$ for the pump beam. Diameters of the two
beams are both approximately $1cm$. Spectrum of the $F = 2 \to F'$
transitions and error signals obtained from the spectrum are
presented in FIG. \ref{Experimental_Result}. To make a comparison,
we present both results obtained from conventional SAS - spectrum
(FIG. \ref{Experimental_Result}(a)) and its error signals (FIG.
\ref{Experimental_Result}(b)) and results obtained from our method -
spectrum (FIG. \ref{Experimental_Result}(c)) and its error signals
(FIG. \ref{Experimental_Result}(d)). In our experiment, crossovers
don't appear on their original places as previous analysis.

\begin{figure}[htbp]
\centering
\includegraphics[width=8.3cm]{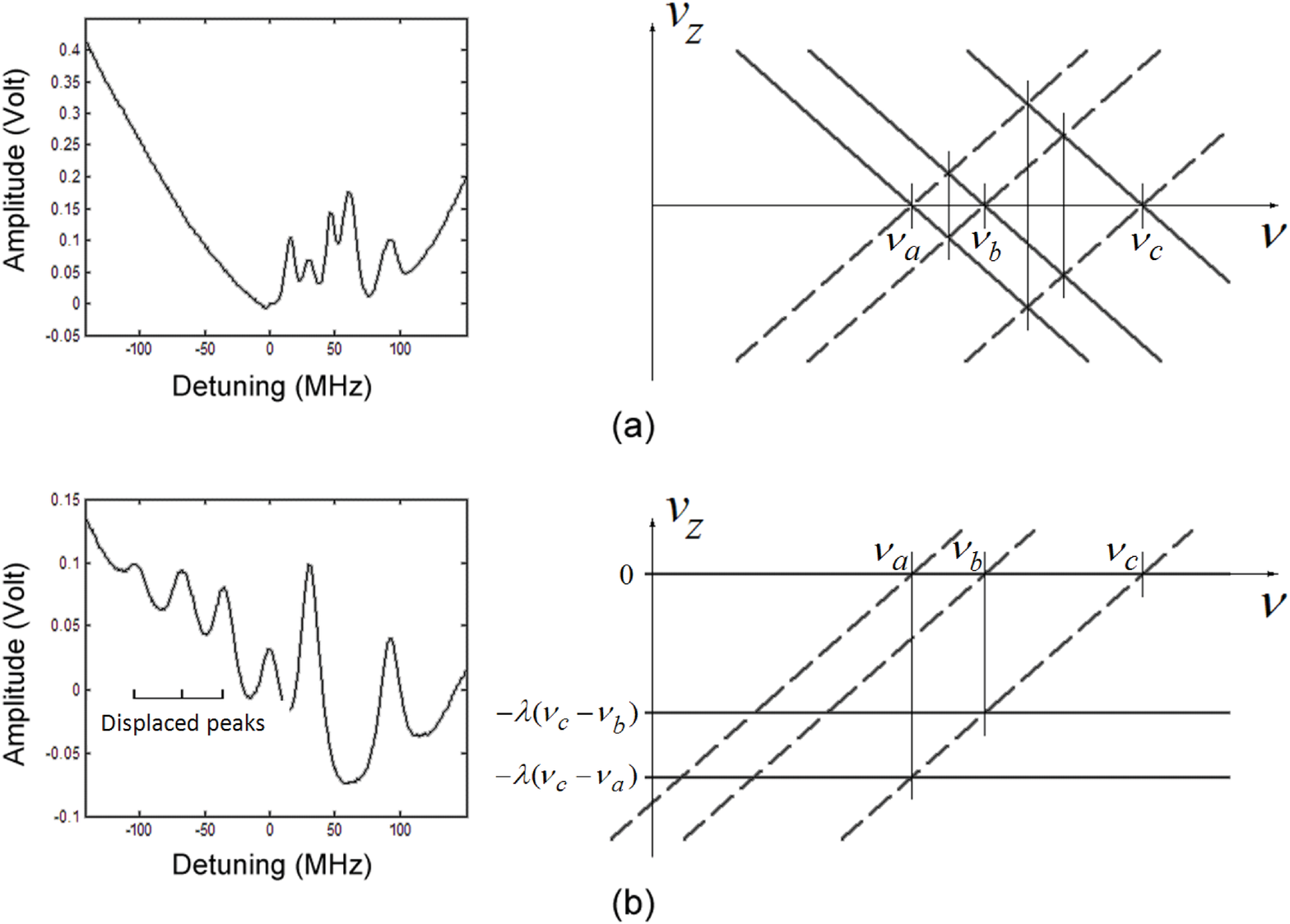}
\caption{Velocity classes of atoms employed by two beams. Dashed
lines represent velocity classes of atoms employed by the probe beam
and solid lines represent those of the pump beam. (a) Conventional
SAS. (b) Our experiment, as its frequency is stabilized, the pump
beam employs atoms of three fixed velocities.}
\label{Detailed_Module_etc}
\end{figure}

Quality of the spectrum is significantly improved. For example, in a
conventional SAS, the peak of $F = 2 \to F' = 1$ transition (the
leftmost one in FIG. \ref{Experimental_Result}(a)) usually looks far
more weak than others\cite{Wieman_Spectrum_a}. Two reasons lead to
this phenomenon. The first reason is that this peak overlaps with
the crossover of $F = 2 \to F' = 1$ and $F = 2 \to F' = 2$ (the
second one in FIG. \ref{Experimental_Result}(a), CO(1,2)). The
second reason is that transition probability for this transition is
relatively smaller. With our method this problem could be overcame
due to the following reasons.

First, the overlap is prevented because there is no crossover in the
vicinity of this peak. We can use $\nu _a$, $\nu _b$ and $\nu _c$ to
represent the three natural resonant frequencies in $F = 2 \to F'$
transitions and plot velocity classes of the atoms employed in FIG.
\ref{Detailed_Module_etc}. Dashed lines represent velocity classes
of atoms employed by the probe beam and solid lines represent those
of the pump beam. The probe beam employs atoms with velocity classes
of ${v_{z,probe}} = \lambda ({\nu _{probe}} - {\nu _{a,b,c}})$
either in conventional SAS or in our experiment. The pump beam
employs atoms with velocity classes of ${v_{z,pump}} = -\lambda
({\nu _{pump}} - {\nu _{a,b,c}})$ in conventional SAS (FIG.
\ref{Detailed_Module_etc}(a)), but because its frequency is fixed to
$\nu_c$ in our experiment, only three velocity classes
(corresponding to states with $F' = 1, 2, 3$) of atoms could be
excited (FIG. \ref{Detailed_Module_etc}(b)). In FIG.
\ref{Detailed_Module_etc}(a), solid lines and dashed lines have
opposite slope because the pump and the probe beam have opposite
directions and always the same frequency. Wherever these two types
of lines having different intercept on the $\nu$ axis 'cross over'
each other, in other words, when two beams having different resonant
frequencies use atoms with the same velocity class, crossover
appears. However, In FIG. \ref{Detailed_Module_etc}(b) frequency of
the pump beam is stabilized so solid lines become horizontal, which
lead to the displacement of crossovers from their original place to
the left of the spectrum.

Second, more atoms are employed to enhance the sensitivity of
detection. When $\nu_{probe} = \nu_a$, in FIG.
\ref{Detailed_Module_etc}(b), the pump beam and the probe beam both
use atoms with ${v_z} = 0$ as well as
\begin{equation}
{v_z} =  - \lambda(\nu_c - \nu_a) \approx - \lambda \times 93MHz
\approx -73m\cdot{s^{ - 1}}, \label{velo_vs_splitting}
\end{equation}
where 93MHz is the separation between $F' = 1$ and $F' =
3$\cite{Rb_data}. However in FIG. \ref{Detailed_Module_etc}(a) only
atoms with ${v_z} = 0$ are used. So roughly twice as many atoms
contribute to the peak, compensating the relatively small transition
probability. Additionally, $\nu_c - \nu_a$ is the energy splitting
of the upper level, the smaller it is, the more likely overlaps with
crossovers take place in conventional SAS. But at the same time
smaller splitting leads to smaller velocity in Eq.
(\ref{velo_vs_splitting}). According to Maxwell velocity
distribution, when this velocity is smaller, larger quantity of
extra available atoms or larger signal could be attained.

Mechanisms discussed above displace peaks of crossovers and enhance
peaks which indicate atom's natural resonant frequencies, making the
spectrum able to show atom's energy structure directly and
distinctly.

Quality of the error signals is also improved. In conventional SAS,
three crossovers lead to partial overlaps of the six peaks,
consequently locking range of each dispersion line shape is
diminished and symmetry of dispersion line shape is destroyed. In
our experiment, first the locking range of every dispersion line
shape is considerably enlarged (FIG. \ref{Experimental_Result}(b),
(d)), especially the first two in FIG. \ref{Experimental_Result}(d)
which once seriously suffered from overlaps with crossovers in
conventional SAS. But merely this is not quite enough to make better
reliability of frequency stabilization. The locking range can be
efficiently used without waste only if symmetry of the line shape is
guaranteed, and this is just our second improvement (FIG.
\ref{Experimental_Result}(d)).

In conclusion, we demonstrated an approach for detecting SAS
different from conventional scheme. We use two independent diode
lasers as the pump beam and the probe beam to move crossovers and to
enhance the sensitivity. The spectrum shows energy levels of atom
more explicitly in comparison with conventional SAS, making it
helpful for observing peaks usually overlapped with crossovers or
peaks corresponding to relatively weak transitions. So this method
could be seem as an experimental aid to discover weak transitions
submerged in crossovers and to observe incompletely analyzed atoms.
Conventional SAS is used to stabilize the pump laser, so usage of
this spectrum and corresponding error signal couldn't lead to better
frequency stability than conventional SAS. Fortunately the quality
improved error signals could provide larger locking range and
symmetry of dispersion line shape, which would significantly enhance
the robustness of frequency stabilization.

\bigskip The authors thank Thibault Vogt's help with English language. This work is supported by the National Fundamental Research
Program of China under Grant No. 2011CB921501, the National Natural
Science Foundation of China under Grant No. 61027016, No.61078026,
No.10874008 and No.10934010.


\pagebreak

\section*{Informational Fourth Page}

\end{document}